\documentclass[12pt]{article}
\usepackage{epsf}
\title{ {\bf
The form factors existing in the $b\rightarrow s g^*$ decay and the possible 
CP violating effects in the noncommutative standard model}}
\author{\vspace{1cm}\\
        {\bf E. O. Iltan}
        \thanks{E-mail address:
        eiltan@heraklit.physics.metu.edu.tr}
 \\
        Physics Department, Middle East Technical University \\
        Ankara, Turkey\\}

\date{}

\begin{document}
\setlength{\baselineskip}{24pt}
\maketitle
\setlength{\baselineskip}{7mm}
\begin{abstract}
We study the form factors appearing in the inclusive decay 
$b\rightarrow s g^*$, in the framework of the noncommutative standard model.
Here $g^*$ denotes the virtual gluon. We get additional structures and 
the corresponding form factors in the noncommutative geometry. We analyse 
the dependencies of the form factors to the parameter $p\,\Theta\, k$ where 
$p$ ($k$) are the four momenta of incoming (outgoing) $b$ quark 
(virtual gluon $g^*$), $\Theta$ is a parameter which measures the
noncommutativity of the geometry. We see that the form factors are weakly 
sensitive to this parameter.  
\end{abstract} 
\thispagestyle{empty}
\newpage
\setcounter{page}{1}
\section{Introduction}
The quantum field theory over noncommutative spaces \cite{Connes} has been 
reached a great interest in recent years . The string theory arguments are 
the reason for the re-motivation of the physics on the noncommutative spaces
\cite{Connes2,Witten}. Noncommutative field theories (NCFT's) are difficult 
to handle since they have non-local structure. Besides this, it has been 
argued that they were sensible field theories and they have been studied
extensively in the literature. The renormalizability of NCFT's in general
have been studied in \cite{Gonzales}. The unitarity in noncommutative     
theories has been discussed in \cite{Gomis}. In \cite{Hewett}, the unitarity
properties of spontaneously broken noncommutative gauge theories have been
examined. The noncommutative quantum electrodynamics (NCQED) and anomalous 
magnetic moments have been studied in \cite{Riad} and a detailed
calculation for the noncommutative electron-photon vertex has been
presented. The noncommutative Yang-Mills theory  has been studied in   
\cite{Krajewski}. In \cite{Hinchliffe} the noncommutative CP violating
effects has been examined at low energies and it was ephasized that CP 
violation due to noncommutative geometry was comparable to the one due to 
the standard model (SM) only, for a noncommutative scale $\Lambda \leq 2\, 
TeV$. Noncommutative SM (NCSM) building has been studied in \cite{Chaichian} 
and recently, the determination of triple neutral gauge boson couplings has 
been done in \cite{Deshpande}.        

In noncommutative geometry the space-time coordinates $x_{\mu}$ are replaced 
by the Hermitian operators $\hat{x}_{\mu}$ where they do not commute
\begin{eqnarray}
[\hat{x}_{\mu},\hat{x}_{\nu}]=i\,\Theta_{\mu\nu} \,\, .
\label{com1}
\end{eqnarray}
Here $\Theta_{\mu\nu}$ is real and antisymmetric tensor. On the other hand
noncommutative field theory is equivalent to the ordinary one except that
the usual product is replaced by the $*$ product 
\begin{eqnarray}
(f*g)(x)=e^{i\,\Theta_{\mu\nu} \,\partial^y_{\mu}\,\partial^z_{\nu}} f(y)\,
g(z)|_{y=z=x}\,.
\label{product}
\end{eqnarray}
The commutation of the Hermitian operators $\hat{x}_{\mu}$ (see eq.
(\ref{com1})) holds with this new product, namely, 
\begin{eqnarray}
[\hat{x}_{\mu},\hat{x}_{\nu}]_*=i\,\Theta_{\mu\nu} \,\, .
\label{com2}
\end{eqnarray}
For constructing the effective low energy theory it is convenient to choose 
the energy scale as $\Lambda=\frac{1}{\sqrt{\Theta}}$ \cite{Hinchliffe}. Here 
the parameter $\Theta$ is taken the average magnitude of the tensor 
$\Theta_{\mu\nu}$. NCSM can be constructed at least up to $O(\Theta)$ by 
replacing the ordinary products by $*$ product. This replacement modifies 
the Feynman rules considerably (see the appendix of \cite{Hinchliffe}).    
  
In our work, we study the form factors appearing in the inclusive decay 
$b\rightarrow s g^*$, in the framework of the NCSM. When the noncommutative
effects are switched on, the form factors due to the SM are modified and 
new structures with the corresponding form factors arise. The noncommutative
effects are at least at the order of $p\,\Theta\, k$ where $p$ ($k$) are the
four momenta of incoming (outgoing) $b$ quark (virtual gluon). Here, we take 
the noncommutative scale $\Lambda=\frac{1}{\sqrt{\Theta}}\leq 1\, Tev$ and 
at these low energies, the problems of unitarity and causality are
supressed. The combination $p\,\Theta\, k$, which appears in the 
expressions are at the order of the magnitude of $10^{-6}-10^{-4}$ for our
process. This is a small number which creates weak noncommutative effects 
in the calculation of form factors. However, these noncommutative effects 
may be stronger for the decays including heavy flavors. On the other hand, 
this parameter is a new source for the CP violation which exists with the 
help of the complex Cabibbo-Kobayashi-Maskawa (CKM) matrix elements in the 
ordinary SM.

The paper is organized as follows:
In Section 2, we present the structures and the form factors appearing in 
the $b\rightarrow s g^*$ decay in the SM, including the non-commutative 
effects. Section 3 is devoted to the analysis of these form factors and our 
discussions.
\section{The form factors existing in the $b\rightarrow s g^*$ decay, in 
the SM including the non-commutative effects} 
\begin{figure}[htb]
\vskip 1.0truein
\centering
\epsfxsize=5.8in
\leavevmode\epsffile{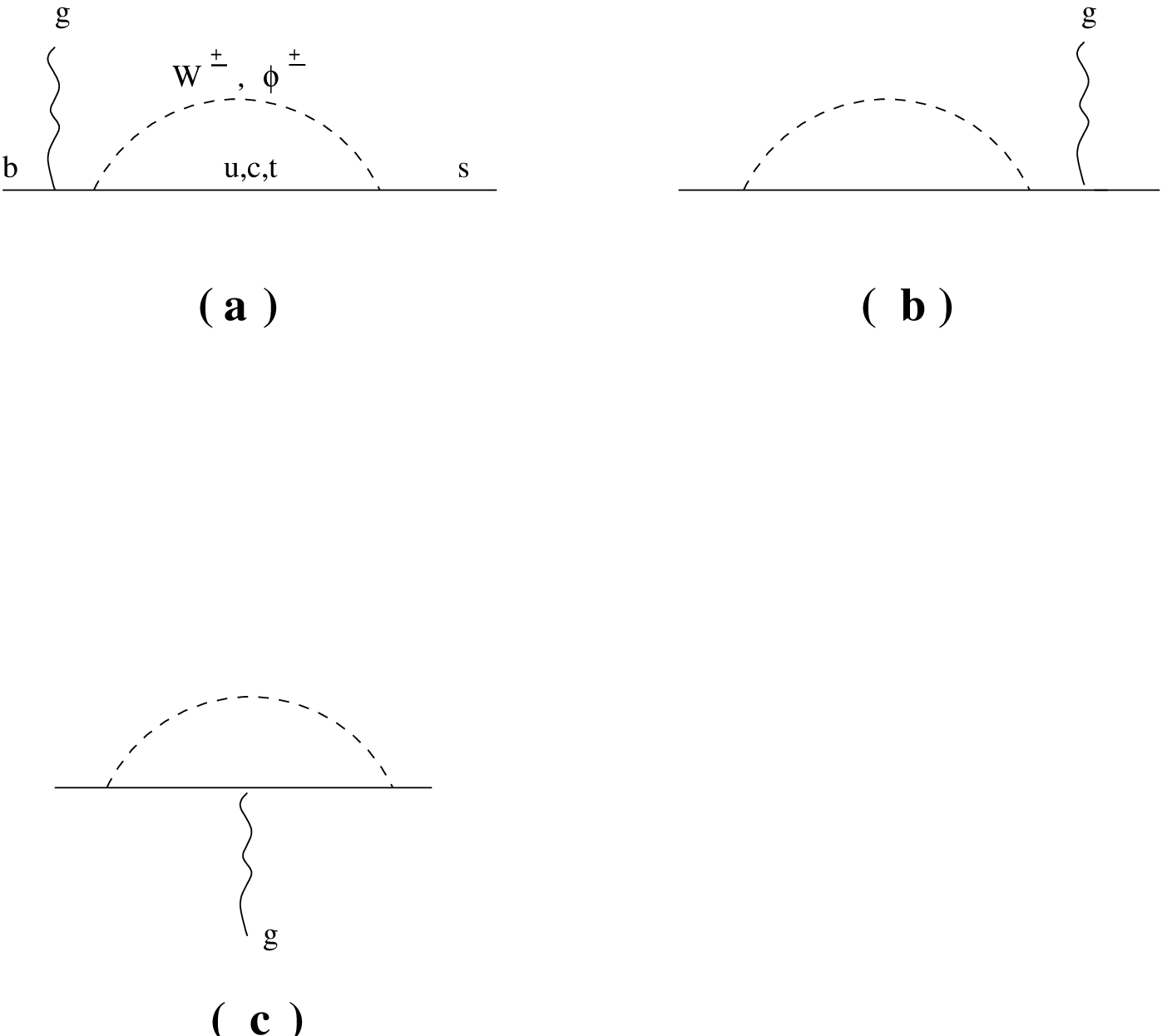}
\vskip 1.0truein
\caption[]{One loop diagrams contribute to $b\rightarrow s g^*$ in the NCSM. 
Wavy lines represent the chromomagnetic field and dashed lines the $W^{\pm}$
and $\phi^{\pm}$ fields.}
\label{fig1}
\end{figure}

In this section, we calculate the form factors of the decay 
$b\rightarrow s g^*$ in the framework of the SM, including the 
noncommutative effects.  As it is well known, $b\rightarrow s g^*$ decay 
is created by flavor changing neutral currents at loop level in the SM. 
The possible interactions at one loop level are self energy and vertex type 
(see Fig.\ref{fig1}). 
At this stage, we use the on-shell renormalization scheme to get rid of the 
divergences appearing in the ordinary  SM and obtain a gauge invariant 
vertex function. Notice that, in this scheme, the self energy diagrams do not 
contribute and only the vertex part survives. When the non-commutative 
effects are switched on, there appears new structures and the corresponding 
form factors, that contain new UV and IR divergences. Use of on-shell 
renormalization scheme helps one to get a gauge invariant vertex function 
however, the form factors appearing due to the noncommutative 
effects still need renormalization. Now, we will present the calculations 
\cite{Riad} to get the form factors for the decay underconsideration.  

The starting point of the calculation is to use the exponential 
representation for the propagators, i.e. the Schwinger parametrization 
\begin{eqnarray}
\frac{i}{p^2-m^2+i\epsilon}=\int_0^{\infty}\, d x_1 
e^{i\,x_1 \,(p^2-m^2+i\epsilon)}
\label{Schwinger}
\end{eqnarray}
and to obtain the denominator of the momentum integral for the vertex 
function, which we call the core integral, 
\begin{eqnarray}
I=-i\, \int \frac{d^d\,q}{(2\pi)^d}\, e^{i\,(q\,z-\frac{1}{2} (p-q)\,
(\widetilde{p'-q})}\, \frac{1}{(q^2-m_W^2)\,\Big( (p-q)^2-m_i^2 \Big) 
\Big( (p'-q)^2-m_i^2 \Big)}\,\, .
\label{int1}
\end{eqnarray}
where  $\widetilde{v_{\mu}}=\Theta_{\mu\nu}\, v^{\nu}$, $p$ ($p'$) is four 
momentum of $b$ ($s$) quark, $p=p'-k$ and $k$ is virtual gluon four momentum. 
Here, the new  factor $e^{-i\,\frac{1}{2} (p-q)\,\widetilde{p'-q}}$ is due 
to the non-commutative geometry and it can be rewritten as 
$e^{-\frac{i}{2}\,p\,\widetilde{p'}}\,\,e^{-\frac{i}{2}\,q\,\widetilde{k}}$. 
The factor $e^{i\,q\,z}$ is introduced to obtain the expressions appearing 
in the numerator of the momentum integral, by differentiation \cite{Zuber}. 
Using the parametrization in eq. (\ref{Schwinger}) and making the momentum 
integration, the integral I in eq. (\ref{int1}) can be written as 
\begin{eqnarray}
I=\int_{0}^{\infty}\, dx_1\,dx_2\,dx_3\, 
\frac{e^{\frac{-i\,(\widetilde{k}+4\,x_2\,p'+4\,x_3\,p-2\,z)^2+
16\,(x_1+x_2+x_3)\,m_W^2\,(x_1+x_i\,(x_2+x_3)-x_b\,x_3)}
{16\,(x_1+x_2+x_3)}}}{(4\pi)^2\,(x_1+x_2+x_3)^2}
\,\, ,
\label{int2}
\end{eqnarray}
where $x_i=\frac{m_i^2}{m_W^2}$, $x_b=\frac{m_b^2}{m_W^2}$. Notice that we
take $m_s=0$ in the expressions. 

Here we will summarize the procedure used in the following:
\begin{itemize}
\item Calculating the the numerator of the integral by differentiating the
momentum integrated core integral  with respect to the auxilary variable $z$
and set $z$ to zero at the end. 
\item Applying the Wick rotation $x_i\rightarrow \frac{x_i}{i}$ and using 
the identity $\int_{0}^{\infty}\,d\rho\, \delta(\rho-(x_1+x_2+x_3) )=1$
\item Making the rescaling $x_i\rightarrow \rho \, x_i$
\item Redefining the core integral by introducing the UV regulator 
$e^{\frac{i}{(x_1+x_2+x_3)\,\Lambda^2}}$ as 
\begin{eqnarray}
I=-\frac{1}{16\,\pi^2}\, e^{-\frac{i}{2}p \widetilde{p'}}\,\int_{0}^{\infty} 
d\rho \int_{0}^{1} dx \int_{0}^{1-x} dy\, 
exp\, [\frac{e_1}{\rho}+e_2 \rho-\frac{i}{2} \widetilde{k}.p\, (1-x)]
\,\, ,
\label{int3}
\end{eqnarray}
with the functions $e_1$ and $e_2$              
\begin{eqnarray}
e_1&=&-\frac{1}{\Lambda_{eff}^2} \,\, , \nonumber \\
e_2&=&-m_W^2\,\Big( x+x_i\,(1-x)-x_b\,(1-x-y)\,(x+s\,y) \Big ) \,\, ,
\label{e12}
\end{eqnarray}
and $\Lambda_{eff}$   
\begin{eqnarray}
\Lambda_{eff}^2=\frac{1}{\frac{1}{\Lambda^2}-\frac{\widetilde{k}^2}{16}} 
\,\, ,
\label{Lambda}
\end{eqnarray}

\end{itemize} 
where $s=\frac{k^2}{m_b^2}$.
With this procedure one obtains all the structures and the corresponding raw 
coefficients due to the decay under consideration. Now, we use the on-shell 
renormalization scheme and extract the nonvanishing structures to get a
gauge invariant result. In this scheme, only the vertex diagram 
(Fig. \ref{fig1}) contributes and the self energy diagrams vanish. Using the 
raw bare vertex function, $\Gamma^{0\, a}_{\mu}$, introducing the counterms 
$\Gamma^{C\, a}_{\mu}$ to satisfy the expression 
\begin{eqnarray}
k^{\mu} \Gamma^{(Raw)\,Ren,\, a}_{\mu}=0 \,\, ,
\label{renorm1}
\end{eqnarray}
where $\Gamma^{(Raw)\,Ren,\, a}_{\mu}$ is 
$\Gamma^{(Raw)\,Ren,\, a}_{\mu}=\Gamma^{0\,a}_{\mu}+\Gamma^{C\,a}_{\mu}$ and  
neglecting the $s$-quark mass we get 
\begin{eqnarray}
\Gamma^{Ren\, a}_{\mu}&=&
\frac{-i\, g^2\,g_s}{32\,m_W^2\,\pi^2}\,\lambda^a \Big( F_1^{Raw} (k^2)\, 
(k_{\mu} \slash \!\!\!{k}-k^2 \gamma_{\mu} ) L + i\,F_2^{Raw} (k^2)\, m_b\,
\sigma_{\mu\nu} \, k^{\nu} R + i\,F_3^{Raw} (k^2) m_b\, \widetilde{k}_{\mu} 
R \nonumber \\ &+& 
i\, F_4^{Raw} (k^2) (\gamma_{\mu} \slash \!\!\!{k} \slash \!\!\!
{\widetilde{k}}-k_{\mu} \slash \!\!\!{\widetilde{k}}) L  + 
F_5^{Raw} (k^2) \widetilde{k}_{\mu} \slash \!\!\!{\widetilde{k}} L  
\label{vertexop}
\end{eqnarray}
where $k_{\mu}$ is the gluon momentum 4-vector and $k^2$ dependent functions 
$F_{1}^{Raw}(k^2)$ and $F_{2}^{Raw}(k^2)$ are proportional to the charge 
radius and dipole form factors. $F_{3}^{Raw}(k^2)$, $F_{4}^{Raw}(k^2)$ and 
$F_{5}^{Raw}(k^2)$ are the new form factors appearing when the noncommutative 
effects are switched on and they read as
\begin{eqnarray}
F_1^{raw} (k^2)&=&\frac{m_W^2}{2}\,\sum_{i=,u,c,t} V_{ib} V^*_{is} 
\int_{0}^{\infty} d\rho \int_{0}^{1} dx \int_{0}^{1-x} dy 
\nonumber \\ & & \!\!\!\!\!\!\!
e^{-\frac{i}{2} p \widetilde{p'}}\,
e^{\frac{e_1}{\rho}+e_2 \rho-\frac{i}{2} \widetilde{k}.p (x-1)} 
\, \Big( x_i\, (x^2-(2-3\,y)\,x+1+2\,y^2-4\,y)+2\,
(x^2+2\,y\,(y-1) \nonumber \\&+& 
x\, (3\,y-2) \Big ) 
\nonumber \,\, ,\\
F_2^{raw} (k^2)&=&\frac{m_W^2}{2}\,\sum_{i=,u,c,t} V_{ib} V^*_{is} 
\int_{0}^{\infty} d\rho \int_{0}^{1}\!\! dx \int_{0}^{1-x}\!\!\!\! dy 
\nonumber \\ & & \!\!\!\!\!\!\!
e^{-\frac{i}{2} p \widetilde{p'}}\,
e^{\frac{e_1}{\rho}+e_2 \rho-\frac{i}{2} \widetilde{k}.p (x-1)} 
\, \Big( x_i+ x^2\, (2+x_i)+ x\, (x_i\, (y-2)+ 2\,y ) \Big) 
\nonumber \,\, ,\\
F_3^{raw} (k^2)&=&\frac{m_W^2}{4}\, \sum_{i=,u,c,t} V_{ib} 
V^*_{is} \int_{0}^{\infty} d\rho 
\int_{0}^{1}\!\! dx \int_{0}^{1-x}\!\!\!\! dy\, 
e^{-\frac{i}{2} p \widetilde{p'}}\,
\frac{e^{\frac{e_1}{\rho}+e_2 \rho-\frac{i}{2} \widetilde{k}.p (x-1)}}{\rho} 
\,(2+x_i)\,(1-x-y) 
\, \, , \nonumber \\
F_4^{raw} (k^2)&=&\frac{m_W^2}{8}\,\sum_{i=,u,c,t} 
V_{ib} V^*_{is} \int_{0}^{\infty} d\rho \int_{0}^{1}\!\! dx 
\int_{0}^{1-x}\!\!\!\! dy\, \frac{e^{-\frac{i}{2} p \widetilde{p'}}\,
e^{\frac{e_1}{\rho}+e_2 \rho-\frac{i}{2} \widetilde{k}.p (x-1)}}{\rho} 
\,(2-\,x_i+x\,(2+x_i))
\, \, , \nonumber \\
F_5^{raw} (k^2)&=&-\frac{m_W^2}{16}\,\sum_{i=,u,c,t} 
V_{ib} V^*_{is} \int_{0}^{\infty} d\rho \int_{0}^{1}\!\! dx 
\int_{0}^{1-x}\!\!\!\! dy\, e^{-\frac{i}{2} p \widetilde{p'}}\,
\frac{e^{\frac{e_1}{\rho}+e_2 \rho-\frac{i}{2} \widetilde{k}.p
(x-1)}}{\rho^2}\,(2+x_i) \,\, ,
\label{Fcofro}
\end{eqnarray}

In the calculation of the coefficients, at first, the $\rho$ integrations 
are taken. These integrations bring the modified Bessel functions of first 
and second type. For the high energy limit, namely $\Lambda^2 \rightarrow 
\infty$, or the low energy limit $k\rightarrow 0$ simultaneously, the 
integration of $F_1^{raw}$  and $F_2^{raw}$ over $\rho$ does not bring any 
divergence. However, the $\rho$ integrations of $F_3^{raw}$  and $F_4^{raw}$  
result in the modified Bessel functions of first type, where the logarithmic 
divercences appear. Here we assume that these divercences can be overcome 
by adding the necessary counter terms, i.e. the NCSM model is renormalizable 
at least at one loop level, similar to NCQED \cite{Hayakawa}. 
The result of the integration of the form factor $F_5^{raw}$ over $\rho$ 
is proportional to the cut-off factor $\Lambda_{eff}^2$. Fortunately, this 
term is irrelevant because of the following reason (see \cite{Riad}). If we 
consider the UV limit, namely, $\frac{1}{\Lambda^2}<< \widetilde{k}^2$ or 
$\Lambda_{eff}^2 \sim \frac{1}{\widetilde{k}^2}$, this term is finite.
However as $\Lambda^2$ tends to zero, we have  
$\frac{1}{\Lambda^2}>> \widetilde{k}^2$ and therefore 
$\Lambda^2 \widetilde{k}^2<<1$. Since the structure due to $F_5^{raw}$ 
contains the term proportional to $\widetilde{k}^2$, this term is irrelevant
in the IR limit. Finally, we end up with the form factors 
$F_1(s)$, $F_2(s)$, $F_3(s)$ and $F_4(s)$:
\begin{eqnarray}
F_1 (s)&=&e^{-\frac{i}{2} p.\widetilde{p'}}\,\sum_{i=,u,c,t} V_{ib} V^*_{is} 
\int_{0}^{1} dx \int_{0}^{1-x} dy 
\nonumber \\ & & e^{-\frac{i}{2} 
p.\widetilde{k} (1-x)}\,\frac{x_i\, (x^2-(2-3\,y)\,x+1+2\,y^2-4\,y)+2\,
(x^2+2\,y\,(y-1) + x\, (3\,y-2)}
{2\,(x_i+ x_b\, y\, (y-1)\,s-x\,(-1+x_i+ x_b\,y\,(1-s) )} 
\nonumber \,\, ,\\
F_2 (s)&=& e^{-\frac{i}{2} p.\widetilde{p'}}\,\sum_{i=,u,c,t} 
V_{ib} V^*_{is} \int_{0}^{1} dx \int_{0}^{1-x}  dy 
\nonumber \\ & & e^{-\frac{i}{2}  p.\widetilde{k}(1-x)}\,
-\frac{x_i+ x^2\, (2+x_i)+ x\, (x_i\, (y-2)+ 2\,y )}
{2\,(x_i+ x_b\, y\, (y-1)\,(s-1)-x\,(-1+x_i+ x_b\,y\,(1-s) )}
\nonumber \,\, ,\\
F_3 (s)&=&e^{-\frac{i}{2} p.\widetilde{p'}}\,\frac{\mbox{EulerGamma}
\, m_b\, m_W^2}{2} \sum_{i=,u,c,t} V_{ib} 
V^*_{is} \int_{0}^{1} dx \int_{0}^{1-x} dy e^{-\frac{i}{2} 
p.\widetilde{k}(1-x)}\,(2+x_i)\,(1-x-y) \, \, , \nonumber \\
F_4 (s)&=&\!\!\!\!-\,e^{-\frac{i}{2} p.\widetilde{p'}}\,\frac{EulerGamma\, 
m_W^2}{4} \sum_{i=,u,c,t} V_{ib} V^*_{is} 
\int_{0}^{1} dx \int_{0}^{1-x} \!\!\! dy e^{-\frac{i}{2} 
p.\widetilde{k} (1-x)}\,(2-\,x_i+x\,(2+x_i)) \, \, . 
\label{Fcof}
\end{eqnarray}
$F_3 (s)$ and $F_4 (s)$, the integration over the parameters $x$ and $y$ 
can be performed easily and we get       
\begin{eqnarray}
F_3 (s)&=& \sum_{i=,u,c,t} V_{ib} V^*_{is}\,\,\mbox{EulerGamma}\,\,m_W^2\,
\frac{e^{-i \,p.\widetilde{p'}}\, \Bigg( 4\, p.\widetilde{p'}+
i\,\Big( (p. \widetilde{p'})^2+8\,(e^{\frac{i \,p.\widetilde{p'}}{2}}-1) 
\Big)\, (2+x_i) \Bigg) }{2 (p.\widetilde{p'})^3}
\nonumber \,\, ,\\
F_4 (s)&=&\sum_{i=,u,c,t} V_{ib} V^*_{is}\,\,\mbox{EulerGamma}\,\,m_W^2\,
\nonumber \\ & &
\frac{e^{-i \,p.\widetilde{p'}}\,\Bigg( 4\, p.\widetilde{p'}\, 
(2\,e^{\frac{i \,p.\widetilde{p'}}{2}}
+x_i)+i\, \Big( (p.\widetilde{p'})^2\, 
(x_i-2)+ 8\,(e^{\frac{i \,p.\widetilde{p'}}{2}}-1)\,(2+x_i) \Big ) \Bigg )}
{2\, (p. \widetilde{p'})^3} \, \, . 
\label{Fcof2}
\end{eqnarray}

For $s > \frac{4\,m_i^2}{m_b^2}$, $i=u,c$, the internal u and c quarks are
on mass-shell and an absorptive part appears in the coefficients related
with the light quark part. Notice that when the non-commutative effects are 
switched off the form factors $F_3(s)$ and $F_4(s)$ disappears and we obtain 
the form factors in the ordinary SM.
\section{Discussion}
This section is devoted to the analysis of the form factors appearing in the
$b\rightarrow s g^*$ process in the framework of the SM including
noncommutative effects. The new parameter existing in this geometry is 
$p\widetilde{k}$ and it is at the order of the magnitude of $10^{-6}-10^{-4}$ 
for the process underconsideration. This factor is also a new source for 
the CP violating effects in addition to the complex CKM matrix elements in 
the SM, $V_{ub}$ in our case. It enters into expressions as an exponential 
and its odd powers in the expansion of the exponential factor bring new CP 
violating effects, even for real CKM matrix elements.  In our work, we study 
$p\widetilde{k}$ and the $sin\delta$ dependence of the real and imaginary 
parts of the form factors. Here, we use the parametrization  
$V_{ub}=e^{i\delta}\, |V_{ub}|$ and $sin\delta$ is proportional to the 
imaginary part of $V_{ub}$, which is the only source of the CP violating 
effects in the commutative SM. Note that we take the numerical value  
$|V_{ub}\,V_{us}|=6\times 10^{-3}$.

\begin{figure}[htb]
\vskip -3.0truein
\centering
\epsfxsize=6.8in
\leavevmode\epsffile{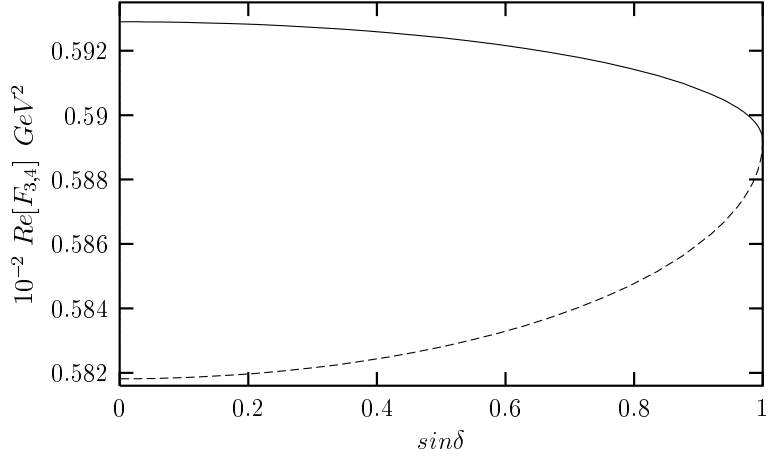}
\vskip -3.0truein
\caption[]{Re[$F_3$] and Re[$F_4$] as a function of $sin\delta$. 
The solid (dashed) line represents Re[$F_3$] (Re[$F_4$]).}
\label{ReF34}
\end{figure}

In Fig. \ref{ReF34} we present the $sin\delta$ dependences of the real parts
of the form factors $F_3$ and $F_4$, for different values of the parameter 
$p\widetilde{k}$, namely $10^{-6}$, $10^{-5}$ and  $10^{-4}$. It is seen
that $Re[F_3]$ and $Re[F_4]$ are not sensitive to $p\widetilde{k}$.
However, there is a weak sensitivity to $sin\delta$. $Re[F_3]$ ($Re[F_4]$) 
decreases (increases) with the increasing values of $sin\delta$. Both form 
factors are almost at the order of the magnitude of $58.5\pm 0.05\, GeV^2$. 

\begin{figure}[htb]
\vskip -3.0truein
\centering
\epsfxsize=6.8in
\leavevmode\epsffile{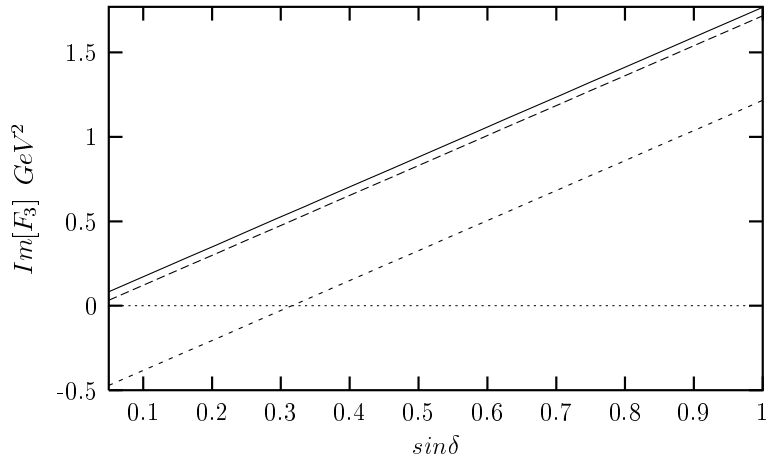}
\vskip -3.0truein
\caption[]{Im[$F_3$] as a function of $sin\delta$. 
The solid (dashed, small dashed) line represents Im[$F_3$]
for $p\widetilde{k}=10^{-6}\, (10^{-5},\,10^{-4})$.}
\label{ImF3}
\end{figure}

\begin{figure}[htb]
\vskip -3.0truein
\centering
\epsfxsize=6.8in
\leavevmode\epsffile{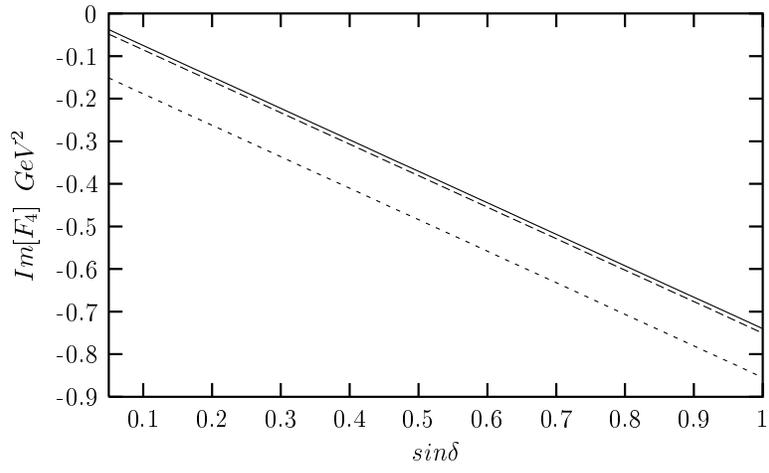}
\vskip -3.0truein
\caption[]{The same as Fig. \ref{ImF3} but for Im[$F_4$].}
\label{ImF4}
\end{figure}

Fig. \ref{ImF3} (\ref{ImF4}) is devoted to the $sin\delta$ dependence of the 
imaginary part of the form factor $F_3$ ($F_4$), for three different values 
of the parameter $p\widetilde{k}$, $10^{-6}$, $10^{-5}$ and  $10^{-4}$. It 
is seen that $Im[F_3]$ ($Im[F_4]$) is sensitive to $p\widetilde{k}$ and it 
increases (decreases) as a magnitude when $p\widetilde{k}$ decreases.   
$Im[F_3]$ can have both signs for large values of $p\widetilde{k}$, depending 
on the parameter $sin\delta$. The sensitivity of $Im[F_3]$ ($Im[F_4]$) to 
the parameter $sin\delta$ is strong. $Im[F_3]$ ($Im[F_4]$) can reach to the 
values $1.7\, GeV^2$ ($-0.9\, GeV^2$)

\begin{figure}[htb]
\vskip -3.0truein
\centering
\epsfxsize=6.8in
\leavevmode\epsffile{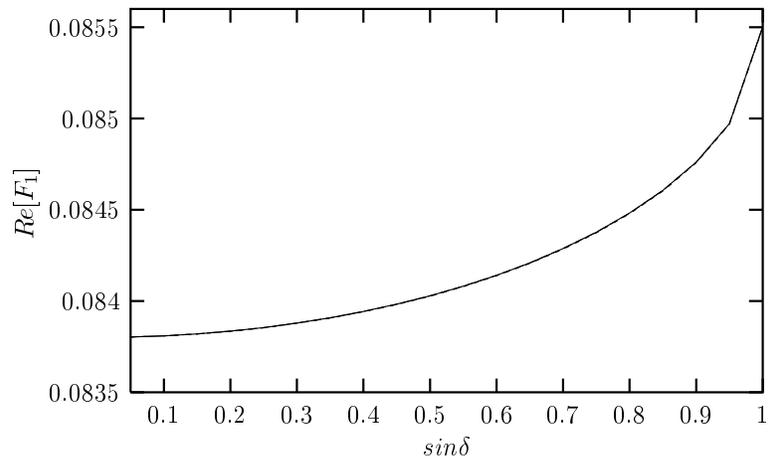}
\vskip -3.0truein
\caption[]{Re[$F_1$] as a function of $sin\delta$.}
\label{ReF1}
\end{figure}

\begin{figure}[htb]
\vskip -3.0truein
\centering
\epsfxsize=6.8in
\leavevmode\epsffile{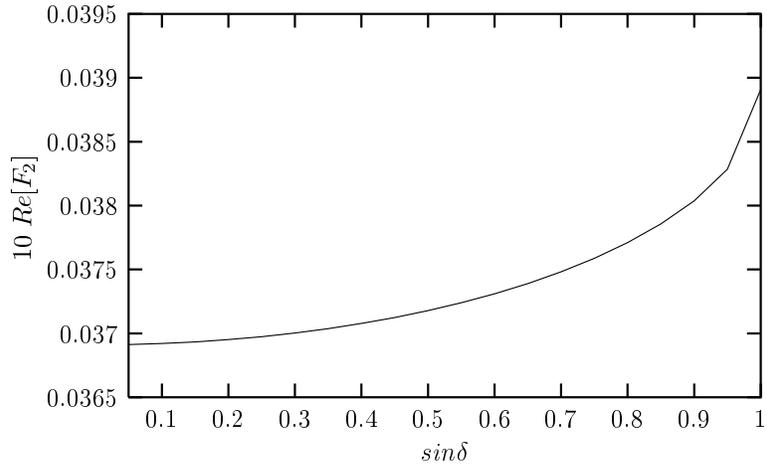}
\vskip -3.0truein
\caption[]{The same as Fig. \ref{ReF1} but for Re[$F_2$].}
\label{ReF2}
\end{figure}

Fig. \ref{ReF1} (\ref{ReF2}) shows the $sin\delta$ dependence of the 
real part of the form factor $F_1$ ($F_2$), for three different values 
of the parameter $p\widetilde{k}$, $10^{-6}$, $10^{-5}$ and  $10^{-4}$.  
$F_1$ ($F_2$) is not sensitive to $p\widetilde{k}$ and weakly sensitive 
to $sin\delta$. It is predicted that its magnitude is $0.0845\pm 0.0005$ 
($0.0038\pm 0.0001$).

\begin{figure}[htb]
\vskip -3.0truein
\centering
\epsfxsize=6.8in
\leavevmode\epsffile{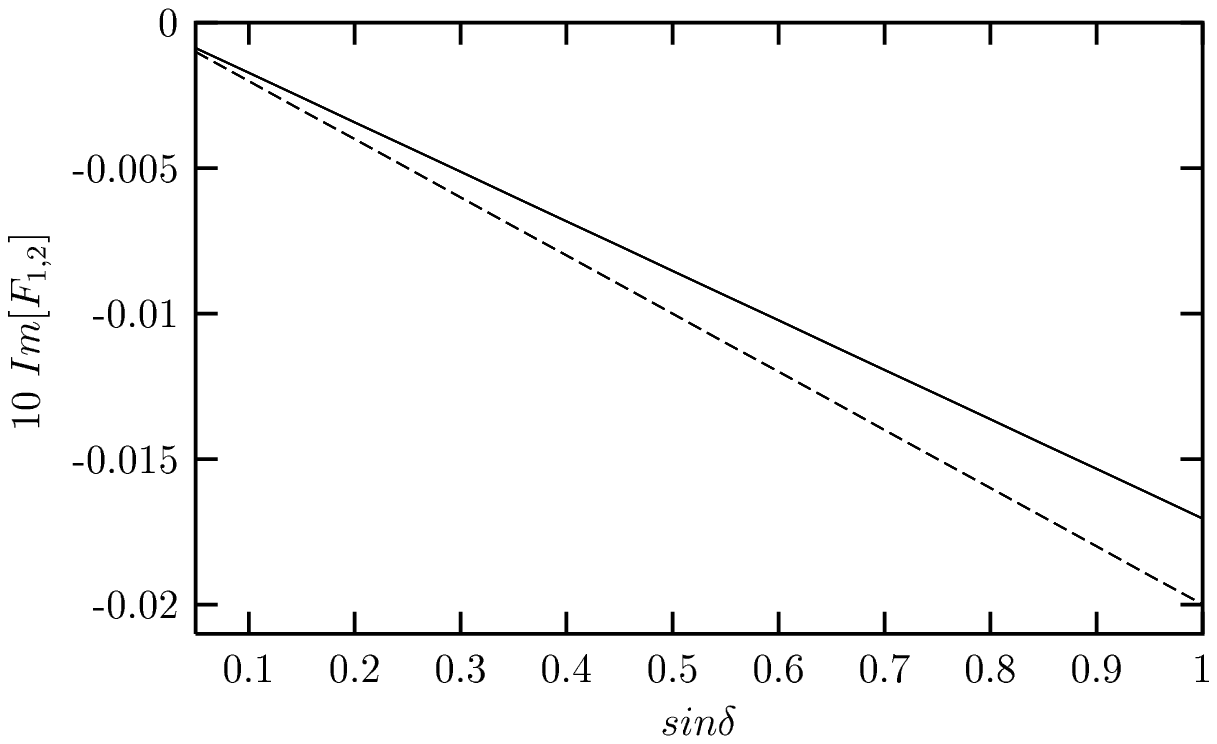}
\vskip -3.0truein
\caption[]{Im[$F_1$] and Im[$F_2$] as a function of $sin\delta$. 
The solid (dashed) line represents Im[$F_1$] (Im[$F_2$]).}
\label{ImF12}
\end{figure}

Finally Fig. \ref{ImF12} is devoted to the $sin\delta$ dependence of the 
imaginary  part of the form factor $F_1$ and $F_2$. 
$Im[F_1]$ and $Im[F_2]$ are not sensitive to $p\widetilde{k}$, however the 
sensitivity to $sin\delta$ not small. Their magnitudes are at the order of 
$0.001\pm 0.001$.

As a summary, there are additional structures and corresponding form factors 
in the non-commutative geometry. The factor $p\widetilde{k}$ is the new
source for the CP violating effects and its magnitude depends on the
process studied. In our case  $p\widetilde{k}$ is small, namely 
$p\widetilde{k}\sim 10^{-5}$ and  the form factors are not so much sensitive 
to this parameter. In the calculation of the CP violation of any decay which 
is based on the process $b\rightarrow s g^*$, the non-commutative effects 
probably weak. However, we belive that these effects would be stronger when 
the processes with heavy flavors have been considered, even in the framework 
of the SM.        
\section{Acknowledgement}
This work was supported by Turkish Academy of Sciences (TUBA/GEBIP).

\newpage
\newpage
\end{document}